\pdfoutput=1
\documentclass[preprint, 12pt]{elsarticle}

\usepackage{my_preamble}
\usepackage{pdfpages}
\usepackage{hyperref}
\journal{Theoretical Computer Science}

\usepackage{lipsum}
\makeatletter
\def\ps@pprintTitle{%
 \let\@oddhead\@empty
 \let\@evenhead\@empty
 \def\@oddfoot{\footnotesize \copyright 2018. \textit{This manuscript version is made available under the \href{http://creativecommons.org/licenses/by-nc-nd/4.0/}{CC-BY-NC-ND 4.0 license.}}
}%
 \let\@evenfoot\@oddfoot}
\makeatother

\begin{document}
\begin{frontmatter}
\title{On the Gap Between Separating Words and Separating Their Reversals}

	\author[farzam]{Farzam Ebrahimnejad}
	\ead{febrahimnejad@ce.sharif.edu}

	\address[farzam]{Department of Computer Engineering, Sharif University of Technology, Tehran, Iran}
	\begin{abstract}
	A deterministic finite automaton (DFA) separates two strings $w$ and $x$ 
if it accepts $w$ and rejects $x$.
The minimum number of states required 
for a DFA to separate $w$ and $x$ is denoted by $\ssep(w,x)$.
The present paper shows that the difference $\left|\ssep(w,x)-\ssep(w^R,x^R)\right|$ 
is unbounded for a binary alphabet; 
here $w^R$ stands for the mirror image of $w$. 
This solves an open problem stated in 
[Demaine, Eisenstat, Shallit, Wilson: Remarks on separating words. 
DCFS 2011. LNCS vol. 6808, pp. 147-157.]
	\end{abstract}

\end{frontmatter}

\bibliographystyle{unsrtnat}
	
\section{Introduction}
	
	In \citeyear{goralvcik1986discerning2}, \citet*{goralvcik1986discerning2} introduced the \textit{separating words problem}. Given two distinct strings $w$ and $x$, we define $\ssep(w,x)$ to be the number of states in the smallest deterministic finite automaton (DFA) that accepts $w$ and rejects $x$ \cite{demaine2011remarks2}. This problem asks for good upper and lower bounds on $$S(n) \coloneqq \max_{w \neq x \land \size{w}, \size{x} \leq n} \ssep(w, x).$$
	\citet{goralvcik1986discerning2} proved $S(n) = o(n)$. Besides, the best known upper bound so far is $O(n^{2/5} \left(\log n\right)^{3/5})$, which was obtained by \citet*{Robson89,Robson96}. A recent paper by \citet*{demaine2011remarks2} surveys the latest results about this problem, and while proving several new theorems, it also introduces three new open problems, all of which have remained unsolved until now. In this paper, we solve the first open problem stated in that paper, which asks whether 
	$$\size{\ssep(w, x) - \ssep(w^R, x^R)}$$
	is bounded or not. We prove that this difference is actually unbounded. In order to do so, in Theorem \ref{th:main} in subsection \ref{subsec:five}, for all positive integers $k$ and $n$, we will construct two strings 	
	$$w = u 0^n v, x = u 0^{n + (2n+1)!} v,$$
	for some $u, v \in \Sets{01, 11}^+ \left( 0^+ \Sets{01, 11}^+\right)^*$, such that $\ssep(w, x) - \ssep(w^R, x^R)$ approaches infinity as $k$ and $n$ approach infinity. As we will later see in Lemma \ref{lem:24} in subsection \ref{subsec:four}, under certain conditions, we can set $u, v$ so that it requires relatively few states to separate $w^R, x^R$. But while preserving these conditions, by using the function $C_n$ and the regular language $G_k$, which we will introduce in subsections \ref{subsec:two} and \ref{subsec:three}, respectively, we can set $u,v$ so that it will require exponentially more states, with respect to $k$, to separate $w$ and $x$. We will see how exactly to do so in the rest of the paper.	
\section{Results}	
	\subsection{Preliminaries}
	We assume the reader is familiar with the basic concepts and terminology of automata theory as in, for example, \cite{Hopcroft}. In this subsection, we present some definitions and notation, and prove a few simple lemmas which will be used in the subsequent subsections.
	
	In this paper, we let $\N$ denote the set of natural numbers, excluding $0$.
	
	\begin{definition}
	We denote a DFA $D$ by a $5$-tuple $(Q_D,\Sigma,\delta_D,q_0,F_D)$, where $Q_D$ is the set of states of $D$, $\Sigma$ is the alphabet that $D$ is defined over, $\delta_D$ is the transition function, $q_0 \in Q_D$ is the start state, and $F_D \subseteq Q_D$ is the set of accept states of $D$.
	\begin{itemize}			
		\item
			For a state $q \in Q_D$ and a string $w \in \Sigma^*$, we define $\delta_D(q, w)$ to be the state in $Q_D$ at which we end if we start reading $w$ from $q$.
			Also, we define $\delta_D(w) \coloneqq \delta_D(q_0,w)$. We say that $D$ accepts $w$ if $\delta_D(w) \in F_D$, and otherwise we say that it rejects $w$.  Moreover, for a subset of states $S \subseteq Q_D$ and a language $L \subseteq \Sigma^*$, we define
			$$\delta_D(S, L) \coloneqq \Set{q' \in Q_D}{\exists q \in S, x \in L : q' = \delta_D(q, x)}.$$
			Finally, we define $\delta_D(q, L) \coloneqq \delta_D(\Sets{q}, L)$.
		\item
			For a positive integer $i$, we define $M_i$ to be the set of all DFAs $E$ defined over $\Sets{0,1,2}$, where $\size{Q_E} \leq i$. Clearly, $M_i$ is finite.
		\item 	
			In this paper, we assume $\Sigma = \Sets{0, 1, 2}$, unless stated otherwise explicitly.
			
		\end{itemize}		
	\end{definition}
		
	\begin{definition}
		Given a DFA $D$ and two distinct strings $w,x \in \Sigma^*$, we say  $D$ \textit{separates} two strings $w$ and $x$, if it accepts $w$ but rejects $x$. Now we can define $\ssep(w,x)$ as the minimum number of states required for a DFA to \separate $w$ and $x$.
		Also, we say that $D$ \textit{distinguishes} $w$ and $x$ if $\delta_D(w) \neq \delta_D(x)$.
	\end{definition}
	Notice that if a DFA \separates two strings, then it must also \distinguish them. The following simple lemma shows that a stronger connection exists between these two definitions.
	\begin{lemma}
		\label{lem:3}
		For any two arbitrary strings $w,x \in \Sigma^*$, if a DFA $D$ \distinguishes $w$ and $x$, then $\ssep(w,x) \le \size{Q_D}$.
	\end{lemma}
	\begin{proof}
	If some DFA $D$ \distinguishes two strings $w,x \in \Sigma^*$, then the DFA with the same set of states and transition function as $D$, but with $\delta_D(w)$ as the only accepting state \separates $w$ and $x$. Therefore we get $\ssep(w, x) \leq \size{Q_D}$.
\end{proof}

	The following lemma shows that adding the same prefix and suffix to two distinct strings will not make it easier to separate them. 
	\begin{lemma}
		\label{lem:4}
		For any four strings $w,x,u,v \in \Sigma^*$ such that $w \neq x$, we have $\ssep(uwv, uxv) \geq \ssep(w, x)$.
	\end{lemma}
	\begin{proof} 
	Let $D$ be a DFA with $\ssep(wv,xv)$ states that \separates $wv$ and $xv$. This DFA must \distinguish $w$ and $x$, so by Lemma \ref{lem:3} we have 
	$$\ssep(w,x) \le |Q_D| = \ssep(wv,xv).$$
	Besides, if some DFA $E$ separates $uwv$ and $uxv$, then the DFA with the same set of states and transitions as $E$ but with $\delta_E(u)$ as the start state \separates $wv$ and $xv$. Hence we have
	$$\ssep(uwv, uxv) \geq \ssep(wv, xv) \geq \ssep(w, x).$$
\end{proof}
	
	The next observation will be used several times throughout this paper, both in Lemma \ref{lem:9} and Theorem \ref{th:main}.
	\begin{proposition}
		\label{prp:five}
		Let $R$ be a regular language. If $x, y \in R\left(0^+R\right)^*$, then $x 0^j y \in R\left(0^+R\right)^*$ for every positive integer $j$.
	\end{proposition}	
	
	Now let us consider the transitions on symbol $0$. The following definition and proposition help us in the proof of Lemma \ref{lem:9} in the next subsection.
	\begin{definition}
		Assume $D$ is a DFA over $\Sets{0,1,2}$. For a state $q \in Q_D$, 
		we say $q$ is in a \textit{zero-cycle}, if there exists some positive integer $i$ such that $\delta_D(q, 0^i)=q$. We call the minimum such $i$ the length of this cycle.

		 Also, for a non-negative integer $i$, we define 
		 $$\zpath_D(q,i) \coloneqq \Set{p=\delta_D(q,0^j)}{0 \le j \le i \text{ and } p \text{ is not in a zero-cycle}}.$$
		  Finally, we denote $\zpath_D(q, \size{Q_D})$ by $\zpath_D(q)$.

	\end{definition}
	
	Notice that if a state $\delta_D(q,0^i)$ is in a \zcycle,
	then for every $j$ with $j>i$, the state $\delta_D(q,0^j)$ is also in a \zcycle.
	Using this fact, we get the following observation.
	
	\begin{proposition}
		\label{prp:one}
		 Let $D = (Q,\Sigma,\delta,q_0,F)$ be a DFA and $i$ be a positive integer. For convenience, we will drop the subscript $D$ from $\zpath_D$. Then
		 \begin{enumerate}[label=(\alph*)]
		 	\item 
		 	\label{st:a}
		 		$\size{\zpath(q,i)} \le i+1$ and $\size{\zpath(q,i)} \le \size{\zpath(q)}$.
		 	\item  
		 	\label{st:b}
		 		If $\delta(q,0^i)$ is not in a \zcycle, then $\size{\zpath(q,i)} = i+1 \le \size{Q}$.
		 	\item 
		 	\label{st:c}
		 		$\size{\zpath(q)} = \size{\zpath(q,i-1)} + |\size{\zpath(\delta(q,0^i))}$.
		 \end{enumerate}
	\end{proposition}
	\begin{proof} 
	\ref{st:a} and \ref{st:b} follow directly from the definition and the fact above.
	To prove \ref{st:c}, notice that $\zpath(q, i - 1) \cap \zpath(\delta(q, 0^i)) = \emptyset$. 
\end{proof}

	\subsection{The Strings $f_n$ and $g_n$, and the Function $C_n$}	
	\label{subsec:two}
	As explained in the Introduction section, our goal is to find some strings $u$ and $v$, so that by setting $w = u 0^n v$ and $x = u 0^{n+(2n+1)!} v$, $\ssep(w,x)$ becomes arbitrarily greater than $\ssep(w^R, x^R)$. The purpose of this subsection is to help us set $u$ and $v$ so that $\ssep(w,x)$ becomes large enough. Actually, it is not hard to show that $\ssep(0^n, 0^{n+(2n+1)!}) = n + 2$. By Lemma \ref{lem:4}, it follows that regardless of what $u$ and $v$ are, the values $\ssep(w,x)$ and $\ssep(w^R,x^R)$ are at least $n+2$. In Lemma \ref{lem:9}, we show that we can set $u$ and $v$ so that $\ssep(w,x) \geq 2n+2$. However, this lemma does not guarantee a low value for $\ssep(w^R, x^R)$, and so Lemma \ref{lem:9} alone does not solve the problem. But still, it plays a crucial role in the proof of Theorem \ref{th:main} in subsection \ref{subsec:five}, and in the next subsections, we will see how to fix this issue.

\begin{definition}
	Since $0^n$ and $0^{n+(2n+1)!}$ are used frequently throughout this paper, from now on, for convenience, we denote them by $f_n$ and $g_n$, respectively.
\end{definition}

	\begin{lemma}
		\label{lem:9}
		For all $n \in \N$ and $w_0 \in \Sigma^+$, there exists $w \in w_0{\left( 0^+ w_0 \right)}^*$ such that $\ssep(wf_nw, wg_nw) \geq 2n + 2$. We denote the $w$ corresponding to $w_0$ by $C_n(w_0)$.
	\end{lemma}
	\begin{proof}
	We run the following algorithm iteratively, while increasing $i$ by $1$ at each step, starting from $i = 1$. While running this algorithm, we preserve the condition that $w_i \in w_0{\left( 0^+ w_0 \right)}^*$. Obviously this condition holds when $i=0$.
	
	In each iteration, if there exists a DFA $D = (Q,\Sigma,\delta,q_0,F) \in M_{2n+1}$ such that $\delta(v, 0^y w_{i-1}) = \delta(v', 0^y w_{i-1})$ for some distinct states $v,v' \in \delta(Q, w_{i-1})$ and some positive integer $y$,  then we set
	$$w_i=w_{i-1} 0^y w_{i-1}.$$
	Otherwise, we set $w_i = w_{i-1}$ and terminate. By the loop condition stated above, we have 	 $w_{i-1} \in w_0{\left( 0^+ w_0 \right)}^*$. Therefore by Proposition \ref{prp:five}, $w_i \in w_0{\left( 0^+ w_0 \right)}^*$ and hence the loop condition holds for $w_i$. Furthermore, let $E$ be an arbitrary DFA in $M_{2n+1}$. Since $w_{i-1}$ is a prefix of $w_i$, 
	if for two states $s, s' \in Q_E$ we have $\delta_E(s, w_{i - 1}) = \delta_E(s', w_{i - 1})$, then $\delta_E(s, w_i) = \delta_E(s', w_i)$. Therefore we have  $\size{\delta_E(Q_E, w_i)} \leq \size{\delta_E(Q_E, w_{i-1})}$.
	Moreover, by the choice of $v$ and $v'$ it follows that 
	$\size{\delta(Q, w_i)} < \size{\delta(Q, w_{i-1})}$. 
	Hence we can write 	$$\sum_{E \in M_{2n+1}}\size{\delta_E(Q_E, w_i)} < \sum_{E \in M_{2n+1}}\size{\delta_E(Q_E, w_{i-1})}.$$
	Thus $\sum_{E \in M_{2n+1}}\size{\delta_E(Q_E, w_i)}$ decreases by at least one in each step, and therefore, this algorithm terminates after a finite  number of iterations. Suppose it terminates after $l$ iterations. We set $w = w_l$.
	
	Now we claim $\ssep(wf_nw, wg_nw) \geq 2n+2$. We prove by backward induction on $t$ that for all $t \geq n$, no DFA in $M_{2n+1}$ can \distinguish $w 0^t w$ and $w 0^t 0^{(2n+1)!} w$. In other words, we will prove by induction on $t$ that for all integers $t \geq n$, DFAs $D \in M_{2n+1}$, and states $q \in \delta(Q, w)$, we have
	$$\delta(q, 0^t w) = \delta(q, 0^t 0^{(2n+1)!} w).$$	
	Base step: Consider $t \geq 2n+1$. Let $D=(Q,\Sigma,\delta,q_0,F)$ be an arbitrary DFA in $M_{2n+1}$. For all states $q \in \delta(Q, w)$, the state 
	$\delta(q, 0^t)$ must be in a \zcycle because otherwise by Proposition \ref{prp:one}, we have
	$$\size{Q} \geq \size{\zpath(q, t)} = t + 1 \geq 2n+2,$$
	which is a contradiction. Since the size of the \zcycle containing $\delta(q, 0^t)$ is at most $\size{Q} \leq 2n+1$, it divides $(2n+1)!$. Thus $\delta(q, 0^t 0^{(2n+1)!}) =  \delta(q, 0^t)$, and hence we have
	$\delta(q, 0^t w) = \delta(q, 0^t 0^{(2n+1)!} w)$.	
	
	Induction step: Consider $n \leq t < 2n + 1$. By the induction hypothesis we know that the claim holds for all $t' > t$. Let $D=(Q,\Sigma,\delta,q_0,F)$ be an arbitrary DFA in $M_{2n+1}$.  
	For convenience, we will drop the subscript $D$ from $\zpath_D$. 
	Pick one of the states $q \in \delta(Q, w)$ maximizing $\size{\zpath(q)}$ amongst all members of $\delta(Q, w)$.  First, we prove the claim for all $p \in \delta(Q, w)$ where $p \neq q$. 
	If $\delta(p, 0^t)$ is in a \zcycle, then by a similar argument as in the base case,
	we obtain $\delta(p, 0^tw) = \delta(p, 0^t 0^{(2n+1)!} w)$, and the proof is complete. Now suppose that $\delta(p, 0^t)$ is not in a \zcycle.
	By Proposition \ref{prp:one}, we have $\size{\zpath(p)} \geq \size{\zpath(p, t)} = t + 1$. Therefore by the choice of $q$, we have $\size{\zpath(q)} \geq \size{\zpath(p)} \geq t + 1$.
	 Hence $\delta(q, 0^t) \in \zpath(q)$, and so it is not in a \zcycle. Thus, we have $\size{\zpath(q, t)} = t + 1$. If $\zpath(q, t) \cap \zpath(p, t) = \emptyset$, then since $\zpath(q, t)$ and $\zpath(p, t)$ are subsets of $Q$, we see that  
	$$\size{Q} \geq \size{\zpath(q, t)} + \size{\zpath(p, t)} = (t + 1) + (t + 1) = 2t + 2 \geq 2n+2,$$
	which is a contradiction. So there exists some $r \in \zpath\left(q, t\right) \cap \zpath\left(p, t\right)$.
	 By definition, there exist $0 \leq a, b \leq t$ such that $\delta(q, 0^a) = \delta(p, 0^b) = r$.     The state $r$ is not in a \zcycle because otherwise, since $t \geq b$, $\delta(p, 0^t)$ should also be in a \zcycle, which contradicts our assumption. Hence by Proposition \ref{prp:one}, we have
	\begin{equation}
		\label{eq:b1}
		\begin{aligned}
			\size{\zpath(q)} &= \size{\zpath(q, a - 1)} + \size{\zpath\left(\delta(q, 0^a)\right)} \\
		&= a + \size{\zpath(r)},
		\end{aligned}
	\end{equation}
	and similarly, we get 
	\begin{equation}
		\label{eq:b2}
		\size{\zpath(p)} = b + \size{\zpath(r)}.
	\end{equation}
	By subtracting equation \ref{eq:b2} from equation \ref{eq:b1}, we obtain
	$$\size{\zpath(q)} - \size{\zpath(p)} = a - b.$$
	But we have $\size{\zpath(q)} \geq \size{\zpath(p)}$. Hence $a \geq b$.
	
	Suppose $a = b$. Then $\delta(q, 0^a) = \delta(p, 0^a)$. So if $a = 0$, then it follows that $p = q$, which contradicts our assumption. 
	Therefore $a > 0$.
	We have $\delta(q, 0^a) = \delta(p, 0^a)$, so $\delta(q, 0^aw) = \delta(p, 0^aw)$. Hence the algorithm could not have terminated, which is a contradiction. Thus we have $a > b$, and so by the induction hypothesis for $(a - b) + t > t$, we have 
	\begin{equation}
		\label{eq:b3}
		\delta(q, 0^{(a-b)+t} w) = \delta(q, 0^{(a-b)+t} 0^{(2n+1)!} w).
	\end{equation}
	But since $b \leq t$, we obtain
	\begin{equation}
		\label{eq:b4}
	\delta(q, 0^{(a-b)+t}) = \delta(q, 0^a 0^{t - b}) = \delta(r, 0^{t-b}) = \delta(p, 0^b 0^{t-b}) = \delta(p, 0^t).
	\end{equation}
	By equations \ref{eq:b3} and \ref{eq:b4}
	we get
	$$\delta(p, 0^t w) = \delta(q, 0^{(a-b)+t} w) = \delta(q, 0^{(a-b)+t} 0^{(2n+1)!} w) = \delta(p, 0^t 0^{(2n+1)!} w),$$
	and therefore the proof is complete for $p$.
	
	It only remains to prove the claim for $q$. Let us write
	$$A = \delta\left(\delta(Q, w) - \Sets{q}, 0^t w\right)$$
	and 
	$$B = \delta\left(\delta(Q, w) - \Sets{q}, 0^t 0^{(2n+1)!} w\right).$$
	We know for any two distinct states $s, s' \in \delta(Q, w)$, we have
	$$\delta(s, 0^t w) \neq \delta(s', 0^t w)$$
	and 
	$$\delta(s, 0^t 0^{(2n+1)!} w) \neq \delta(s', 0^t 0^{(2n+1)!} w)$$
	because otherwise the algorithm could not have terminated, which is a contradiction. So $\size{A} = \size{B} = \size{\delta(Q, w)} - 1$. But we proved the induction step for all members of $\delta(Q, w)$ except $q$. Hence for all states $s \in \delta(Q, w) - \Sets{q}$ we have $\delta(s, 0^t w) = \delta(s, 0^t 0^{(2n+1)!} w)$. Therefore $A = B$. 
	Let us write $e = \delta(q, 0^t w)$ and $e' = \delta(q, 0^t 0^{(2n+1)!} w)$. Since $w$ is a suffix of both $0^t w$ and $0^t 0^{(2n+1)!} w$, by definition we have $e, e' \in \delta(Q, w)$. Also, since the algorithm has terminated, we get $e \notin A$ and $e' \notin B$. Consequently we have
	$$e \in \delta(Q, w) - A,$$
	and
	$$e' \in \delta(Q, w) - B = \delta(Q, w) - A.$$
	But since $w$ is a suffix of $0^tw$, we have $A \subseteq \delta(Q, w)$. So
	$$\size{\delta(Q, w) - A} = \size{\delta(Q, w)} - \left( \size{\delta(Q, w)} - 1 \right) = 1.$$
	Therefore $e = e'$ and the proof is complete.
\end{proof}

	\subsection{The Regular Language $G_k$}
	\label{subsec:three}
	In this subsection, we introduce the regular language $G_k \subseteq \Sets{1,2}^*$, which has some interesting characteristics. For all $k \in \N$,
	there exists a DFA with $O(k)$ states that accepts $G_k^R$, while no DFA with less than $2^k$ states accepts $G_k$. 
	Similar regular languages that also have these two characteristics
 	have been defined before \cite{reverseGaoKY12a, reverseJiraskova08, reverseSebej10} but are not quite appropriate for our purposes. Another characteristic of $G_k$ is that, as proven later in Lemma \ref{lem:17}, there exists $z_k \in G_k$ such that if a DFA with less than $2^k$ states accepts $z_k$, then it should also accept some string in $\Sets{1,2}^* - G_k$. This, together with Lemma \ref{lem:9}, helps us construct the desired strings in Theorem \ref{th:main}. Recall that $\N$ denotes the set of positive integers.

	\begin{definition}
		 For every positive integer $k$, we define languages $L_k$ and $G_k$ over $\Sets{1,2}$ as follows:
		\begin{equation*}
		 	\begin{aligned}
		 		L_k \coloneqq& \Set{1^{2i}2}{i \in \N \wedge i \leq k}
				\\
				\cup& \left\{1^{i_1}21^{i_2}2   \cdots 21^{i_{s - 1}}21^{i_s}2\right. &&\mid{} s, i_1, i_2, \ldots, i_s \in  \N \\
				&  &&\wedge{} i_1 + i_2 + \cdots + i_s = 2k + 1 \\
				& &&\wedge{} i_1, i_2, \ldots, i_{s - 1} \equiv 0  \left. \pmod{2} \right\}.
		 	\end{aligned}
		\end{equation*}
		Finally, we define $G_k \coloneqq L_k^*$
	\end{definition}
	
	\begin{lemma}
		\label{lem:11}
		For all $u, v \in \Sigma^*$, we have $u1^{2k+1}2v \in G_k$ if and only if $u,v \in G_k$.	
	\end{lemma}
	\begin{proof}
	We can easily observe that if $x 1^{2k+1}2 y \in L_k$, then $x = y = \epsilon$. Thus it follows that if $u 1^{2k+1}2 v \in L_k^* = G_k$, then both $u$ and $v$ should also be in $G_k$.
	
	For the other direction, obviously we have $1^{2k+1}2 \in L_k$. Therefore by definition, if $u, v \in G_k$ then $u 1^{2k+1}2 v \in G_k$.
\end{proof}
	
	\begin{definition}
		For a regular language $L \subseteq \Sigma^*$, we define $\stc(L)$, or the state complexity of $L$, to be the minimum number of states required for a DFA to accept $L$. This concept has been studied for a long time; see, for example, \cite{maslov1970estimates, YuZS92, YuZS94}.
	\end{definition}

	\begin{lemma}
		\label{lem:13}
		For all integers $k \in \N$, we have $\stc(G_k) \geq 2^k$.
	\end{lemma}
	\begin{proof}
	Let $E$ be the set of all positive even numbers less than $2k+1$ and $\powerset{E}$ be the set of all subsets of $E$. We define the function $r: \powerset{E} \rightarrow \Sets{1,2}^*$ as follows:
	
	For the empty set, we define $r(\emptyset) \coloneqq \epsilon$. Now consider an arbitrary non-empty subset of $E$, such as
	$$S = \Sets{a_1, a_2, \ldots, a_m}.$$
	Without loss of generality, assume $a_1 < a_2 < \cdots < a_m$. We define
	$$r(S) \coloneqq 1^{a_m-a_{m-1}} 2 1^{a_{m-1}-a_{m-2}} 2 \cdots 2 1^{a_2-a_1} 2 1^{a_1} 2.$$
	
	Let $1 \leq i < 2k + 1$ be an odd number and $S \subseteq E$. We claim $r(S)1^i2 \in G_k$ if and only if $2k+1-i \in S$.  If $w=r(S)1^i2 \in G_k$, then by definition $x \in G_k$ and  $y \in L_k$ exist such that $w = xy$. But $i$ is an odd number and $1^i 2$ is a suffix of $y$. Therefore by definition, $1 \leq p \leq m$ and  $b_1, b_2, \ldots, b_p \in \N$ exist such that 
	$$b_p + b_{p - 1} + \cdots + b_1 + i = 2k + 1,$$
	and
	$$y = 1^{b_p} 2 1^{b_{p - 1}} 2 \cdots 1^{b_1} 2 1^i 2.$$
	It follows that
		$$b_1 = a_1, b_2 = a_2 - a_1, \ldots, b_p = a_p - a_{p - 1}.$$
	Thus
		$$a_p = b_1 + \cdots + b_p = 2k + 1 - i.$$
	So $2k + 1 - i \in S$.
	
	For the other direction, suppose $2k + 1 - i \in S$. Then $a_j \in S$ exists such that $a_j = 2k + 1 - i$. All members of $S$ are even numbers less than $2k+1$. Hence, by definition we have
	$$ 1^{a_m-a_{m-1}} 2, 1^{a_{m-1}-a_{m-2}} 2, \ldots, 1^{a_{j+1}-a_j} 2 \in L_k.$$
	Moreover, we have
	$$ i + (a_1 + \sum_{t=2}^{j}{a_t - a_{t-1}})
	 = (2k + 1 - a_j) + a_j = 2k + 1.$$
	Hence
	$$ 1^{a_j - a_{j - 1}} 2 \cdots 1^{a_2 - a_1} 2 1^{a_1} 2 1^{i} 2 \in L_k.$$
	Therefore $r(S) 1^i 2 \in G_k = L_k^*$.


Now consider the family of strings $\Set{r(S)}{S \in \powerset{E}}$ of size $2^k$.
Let $S$ and $S'$ be two distinct sets in this family. Then, without loss of generality, there is an integer $c$ with $c \in S \setminus S'$.
Therefore, by our claim above, we have $r(S)1^{2k+1-c} \in G_k$ while $r(S')1^{2k+1-c} \notin G_k$.
It follows that $\stc(G_k) \geq 2^k$.

\end{proof}
		
	\begin{lemma}
		\label{lem:14}
		For all integers $k \in \N$, we have $\stc(G_k^R) \leq 5k + 3$.
	\end{lemma}
	\begin{proof}
	It suffices to prove there exists a DFA $D \in M_{5k+3}$ such that $L(D) = G_k^R$. We define $D=(Q, \Sigma, \delta, p_{2k+1}, F)$ as follows:
	
	We set 
	$$Q = \Set{p_i}{0 \le i \le 2k+1} \cup \Set{r_i}{2\le i \le 2k+1} \cup \Set{r'_{2i-1}}{1 \le i \le k} \cup \Sets{d}$$
	Additionally, we specify the following rules for the transition function:
	\begin{itemize}
		\item
	$\delta(p_i, 1) = p_{i+1}$ $(0 \leq i \leq 2k)$,
		\item
		$\delta(r_i, 1) = r_{i+1}$ $(2 \leq i \leq 2k)$,    
		\item  $\delta(r_{2i-1}, 2) = r'_{2i - 1}$ $(2 \leq i \leq k)$,
		\item
	    $\delta(r'_{2i-1}, 1) = r_{2i}$ $(1 \leq i \leq k)$, 
	    
		\item
		$ \delta(p_i,2)=\begin{cases}
		p_0, & \text{if } 2 \le i \le 2k \text{ and } i \text{ is even;} \\
		r'_i, & \text{if } 1\le i \le 2k \text{ and } i \text{ is odd, }
		\end{cases} $


		\item
		$\delta(p_{2k+1},2) = \delta(r_{2k+1},2) = p_0$,
	\end{itemize}
	and all the remaining transitions go to the dead state $d$. The DFA $D$ is shown in Figure \ref{fig:14}.
	
		 Finally, we set
	$$F = \Set{p_{2i}}{1 \le i \le k} \cup \Sets{p_{2k+1}, r_{2k+1}}.$$

	\begin{figure}
\centering

\includegraphics[width=0.9\textwidth]{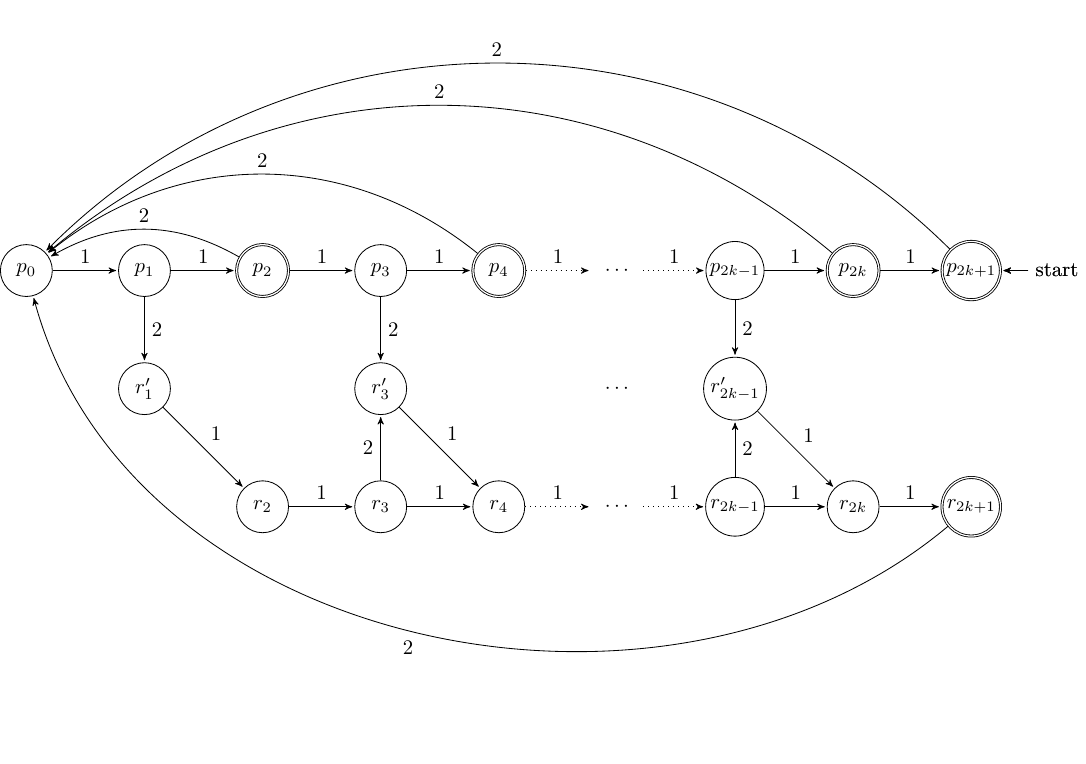}
\caption{The DFA $D$ which is explained in Lemma \ref{lem:14}. The reject state $d$ is not shown.}
  \label{fig:14}
\end{figure}
	It is not hard to verify that $\delta(F, L_k^R) \subseteq F$, and hence $\delta(F, \left(L_k^R\right)^*) =\delta(F, G_k^R) \subseteq F$. It is also easy to show that $\delta(F, \Sigma^* - G_k^R) \cap F = \emptyset$. Thus since $p_{2k+1} \in F$, we obtain $L(D) = G_k^R$. Therefore we have $\stc(G_k^R) \leq \size{Q} = 5k + 3$.

\end{proof}

	\begin{definition}
		For $w \in \Sigma^*$ and a language $L$ over $\Sigma$, we define $\lsep(w, L)$ as the minimum number of states of a DFA that accepts $w$ and rejects all $x \in L$.
	\end{definition}
	
	\begin{definition}
		Since the set $\Sets{1,2}^* - G_k$
		 is referred to several times in the rest of this paper, for simplicity, we will denote it by $H_k$.
	\end{definition}

	\begin{lemma}
		\label{lem:17}
		There exists $z_k \in (G_k - \Sets{\epsilon})$ such that $\lsep(z_k, H_k) \geq 2^k$.
	\end{lemma}
	\begin{proof}
	At the beginning, we set $w_0=\epsilon$, $U_0 = M_{2^k-1}$, and $V_0 = \emptyset$. We preserve the following conditions for all $j \geq 0$:
	\begin{enumerate}
		\item $V_j \cup U_j = M_{2^k-1}$;
			\label{dnoc:first}
		\item
			\label{dnoc:third}
			$w_j \in G_k$;
		\item For all DFAs $D \in V_j$, there exists some $r \in H_k$ such that $D$ \ndistinguishes $r$ and $w_j$.
			\label{dnoc:second}

	\end{enumerate} 
	Obviously these conditions hold for $j = 0$. 

	Now we run the following algorithm iteratively, while increasing $i$ by $1$ at each step, starting with $i=1$:
	
	In each iteration, if there exists a DFA $D =(Q,\Sigma,\delta,q_0,F) \in U_{i - 1}$, and strings $x \in G_k$ and $y \in H_k$ such that $\delta(w_{i-1} 1^{2k+1} 2 x) =  \delta(w_{i-1} 1^{2k+1} 2 y)$, then we set $w_i = w_{i-1}  1^{2k+1} 2 x$, $U_i = U_{i-1} - \Sets{D}$, and $V_i = V_{i-1} \cup \Sets{D}$.
	Otherwise, we terminate by setting $w_i=w_{i-1}$, $U_i = U_{i-1}$ and $V_i = V_{i-1}$.

	Obviously, Condition \ref{dnoc:first} holds for $j=i$. Moreover, by Condition \ref{dnoc:third} for $j=i-1$, we have $w_{i-1} \in G_k$. Therefore by Lemma \ref{lem:11}, we have $w_i \in G_k$, and hence Condition \ref{dnoc:third} holds for $j=i$.
	
	Furthermore, by Condition \ref{dnoc:second} for $j=i - 1$, for all DFAs $E \in V_{i - 1}$, there exists $r \in H_k$ such that $\delta_E(r) = \delta_E(w_{i-1})$. Hence 
	we have 
	$$\delta_E(r1^{2k+1}2x) = \delta_E(w_{i-1}1^{2k+1}2x) = \delta_E(w_i).$$
	But by Lemma \ref{lem:11}, we obtain  $r1^{2k+1}2x \in H_k$. Thus Condition \ref{dnoc:second} for $j = i$ holds for all members of $V_{i - 1} = V_i - \Sets{D}$.
	It only remains to prove that it also holds for $D$. We have 
	$$ \delta(w_i) = \delta(w_{i-1}1^{2k+1}2x) 
	  = \delta(w_{i-1}1^{2k+1}2y).$$
	 But by Lemma \ref{lem:11}, we get $w_{i-1}1^{2k+1}2y \in H_k$. Hence Condition \ref{dnoc:second} holds for $D$. Therefore Condition \ref{dnoc:second} holds for $j = i$.
		
		This algorithm terminates after a finite number of iterations because $|U_i|$ decreases by $1$ at each step (except the last one).  Suppose it terminates after $t$ iterations. We claim $U_t = \emptyset$. Otherwise there exists some DFA $B \in U_t$. Let $B'$ be the DFA with the same set of states and transition function as $B$, but with $\delta_B(w_{t-1} 1^{2k+1} 2)$ as the start state and with $\delta_B(\delta_B(w_{t-1} 1^{2k+1} 2), G_k)$ as the set of accepting states. By definition, we have $G_k \subseteq L(B')$. Furthermore, $B'$ cannot accept any string $w \notin G_k$ because otherwise 
		$$\delta_B(\delta_B(w_{t-1} 1^{2k+1} 2), G_k) \cap \delta_B(\delta_B(w_{t-1} 1^{2k+1} 2), H_k) \neq \emptyset,$$
		 	and therefore the algorithm could not have terminated, which is a contradiction. Hence $L(B')=G_k$. So by Lemma \ref{lem:13} we have $B' \notin M_{2^k-1}$, which contradicts $\size{Q_{B'}} = \size{Q_B} \leq 2^k-1$.
		
		Thus by Condition \ref{dnoc:first} it follows that $V_t = M_{2^k-1}$. By Condition \ref{dnoc:second}, for all DFAs $D \in M_{2^k - 1}$, if $D$ accepts $w_t$, then it also accepts some string in $H_k$. Hence we obtain $\lsep(w_t, H_k) \geq 2^k$. By Condition \ref{dnoc:third}, we have $w_t \in G_k$. Since $V_t$ is not empty, we obtain that the algorithm has terminated after a positive number of iterations. Furthermore, for $1 \leq i \leq t$, the string $w_i$ starts with $1^{2k+1}2$.  Hence $w_t$ is not empty, and so we have $w_t \in (G_k - \Sets{\epsilon})$. Therefore we can set $z_k \coloneqq w_t$.
\end{proof}
	
	\begin{definition}
		The set $H_k \cup \Sets{z_k}$
		 is referred to several times in the rest of this paper. So, for simplicity, we will denote it by $H'_k$.
	\end{definition}
	
	\begin{remark}
		\label{rem:spider}
		For any two DFAs $D \in M_i$ and $D' \in M_j$, some DFA $E \in M_{i \times j}$ exists such that $L(E) = L(D) \cap L(D')$.	
	\end{remark}

	\begin{lemma}
		\label{lem:19}
		For every two DFAs $D, D' \in M_{2^{k/2}-1}$, and every string $w \in \Sigma^*$, there exists $x \in H_k$ such that $\delta_D(w z_k) = \delta_D(w x)$ and $\delta_{D'}(w z_k) = \delta_{D'}(w x)$.
	\end{lemma}
	\begin{proof}
	Let $D = (Q, \Sigma, \delta, q_0, F)$ and $D' = (Q', \Sigma, \delta', q'_0, F')$. 
	We set $E$ to be the same DFA as $D$ but with $\delta(w)$ as the start state, and with $\delta(w z_k)$ as the only accept state. Similarly, we set $E'$ to be the same DFA as $D'$ but with $\delta'(w)$ as the start state, and with $\delta'(w z_k)$ as the only accept state. 
	Obviously, $z_k \in L(E) \cap L(E')$. 
	By Lemma \ref{lem:17}, $L(E) \cap H_k$ and $L(E') \cap H_k$  are not empty.
	 We further claim that their intersection, $L(E) \cap L(E') \cap H_k$, is also not empty. 	 Otherwise, by Remark \ref{rem:spider}, some DFA $F$ with at most 
	$$(2^{k/2} - 1)(2^{k/2} - 1) < 2^k$$
	states exists such that $L(F) = L(E) \cap L(E')$. If $L(E) \cap L(E') \cap H_k = \emptyset$, then we obtain $L(F) \cap H_k = \emptyset$. But $z_k \in L(F)$. Hence $F$ accepts $z_k$ but rejects every string
	 in $H_k$, and therefore, by Lemma \ref{lem:17}, we have $\size{Q_F} \geq 2^k$, which is a contradiction. Hence there exists some
	 $$x \in L(E) \cap L(E') \cap H_k,$$
	or equivalently, both $E$ and $E'$ accept some $x \in H_k$. Furthermore, by the construction of $E$ and $E'$, we obtain $\delta(w z_k) = \delta(w x)$ and $\delta'(w z_k) = \delta'(w x)$,
	and therefore the proof is complete.
\end{proof}

	\begin{lemma}
		\label{lem:20}
		Let $w \in H'_k \left(0^+ H'_k \right)^*$. For any two DFAs $D, D' \in M_{2^{k/2}-1}$, there exists some $w' \in H_k \left(0^+ H_k \right)^*$ such that $\delta_D(w) = \delta_D(w')$ and $\delta_{D'}(w) = \delta_{D'}(w')$, or in other words, neither $D$ nor $D'$ \distinguishes $w$ and $w'$.
	\end{lemma}
	\begin{proof}
	We have $w \in H'_k \left(0^+ H'_k \right)^*$. So it can be expressed as 
	$$w=u_1 0^{i_1} \cdots 
u_{l-1} 0^{i_{l-1}} u_l 0^{i_l} u_{l+1},$$
	where $i_1, \ldots, i_l \in \N$ and $u_1, \ldots, u_{l+1} \in H'_k$. For $1 \leq j \leq l$, let us write
	$$w_j=u_1 0^{i_1}  \cdots u_{j-1} 0^{i_{j-1}} u_{j} 0^{i_j}.$$
	For simplicity, we also set $i_0 = 0$ and $w_0 = \epsilon$. Now for $1 \leq j \leq l + 1$, we define the strings $u'_j \in H_k$ as follows: If $u_j \neq z_k$, then we set $u'_j = u_i$. Otherwise,
	let $D = (Q, \Sigma, \delta, q_0, F)$ and $D' = (Q', \Sigma, \delta', q'_0, F')$. By Lemma \ref{lem:19}, it follows that there exists $x \in H_k$ such that $\delta(w_{j-1} z_k) = \delta(w_{j-1} x)$ and $\delta'(w_{j-1} z_k) = \delta'(w_{j-1} x)$.
		We set $u'_j = x$. In either of the cases, clearly we have $u'_j \in H_k$.
	Now let us write $$w'=u'_1 0^{i_1} \cdots u'_l 0^{i_l} u'_{l+1}.$$
	We claim $\delta(w) = \delta(w')$ and $\delta'(w) = \delta'(w')$. Let us set $x_0 = x'_0  = \epsilon$. Moreover, for $1 \leq j \leq l + 1$, we set
	$$x_j = u_1 0^{i_1}  \cdots u_{j-1} 0^{i_{j-1}} u_j,$$
	and
	$$x'_j = u'_1 0^{i_1}  \cdots u'_{j-1} 0^{i_{j-1}} u'_j.$$
	We prove by induction that for $0 \leq j \leq l + 1$, we have $\delta(x_j) = \delta(x'_j)$ and $\delta'(x_j) = \delta'(x'_j)$.  The base step is obvious for $j = 0$. For $j \geq 1$, if $u_j \neq z_k$, then we have $u'_j = u_j$, and 
	so we can obtain the claim.
	Otherwise, by the induction hypothesis we have $\delta(x_{j-1}) = \delta(x'_{j-1})$. 
	By the choice of $u'_{j-1}$ we have
	\begin{equation*}
		\begin{aligned}
			\delta(x_j) &= \delta(w_{j-1} z_k) 
			 = \delta(w_{j-1} u'_j) &&=  \delta(x_{j-1} 0^{i_{j-1}} u'_j) \\
			 & &&= \delta(x'_{j-1} 0^{i_j-1} u'_j) = \delta(x'_j),
		\end{aligned}
	\end{equation*}
	and the proof of the claim is complete. Similarly, we can prove $\delta'(x_j) = \delta'(x'_j)$ for all $0 \leq j \leq l + 1$. Hence we obtain 
	$$\delta(w) = \delta(x_{l+1}) = \delta(x'_{l+1}) = \delta(w')$$
	and 
	$$\delta'(w) = \delta'(x_{l+1}) = \delta'(x'_{l+1}) = \delta'(w').$$
	Besides, for all $1 \leq j \leq l + 1$, we have $u'_j \in H_k$. Therefore it follows that $w' \in H_k \left(0^+ H_k \right)^*$, and hence the proof is complete.

\end{proof}
	
	\begin{proposition}
		\label{prp:four}
		Let $D$ be a DFA in $M_{2^{k/2}-1}$, $q, q' \in Q_D$, and $w \in H'_k \left(0^+ H'_k \right)^*$. There exists some $w' \in H_k \left(0^+ H_k \right)^*$ such that $\delta_D(q, w) = \delta_D(q, w')$ and $\delta_D(q', w) = \delta_D(q', w')$. 
	\end{proposition}
	\begin{proof}
	We define two new DFAs $E$ and $E'$, having the same set of states and transition function as $D$, but with $q$ and $q'$ as their starting states, respectively. 		The proposition follows directly from applying Lemma \ref{lem:20} to $E,E'$ and $w$. \end{proof}

	\subsection{Mapping $\Sets{0,1,2}^*$ to $\Sets{0,1}^*$}
	\label{subsec:four}
	The previous lemmas may help us to construct two strings in $\Sigma^*=\Sets{0,1,2}^*$ with our desired characteristics. But our goal is to prove our result for an alphabet of size $2$. To be able to construct the intended strings over $\Sets{0,1}$, in this subsection we introduce the function $\tr$ that maps strings in $\Sigma^*$ to strings in $\Sets{0,1}^*$, while preserving some of our desired characteristics in them.
	
	\begin{definition}
		For a string $w \in \Sigma^*$, we define $\tr(w)$ to be the string obtained from $w$ by replacing all occurrences of $1$ by $11$ and all occurrences of $2$ by $10$. Clearly we have $\tr(w) \in \Sets{0,1}^*$		
	\end{definition}
	
	The following lemma shows that when two strings are mapped under $\tr^R$, separating them would be at least as hard as separating the original ones.
	\begin{lemma}
		\label{lem:23}
		For all pairs of distinct strings $w,x \in \Sigma^*$, we have 
		$$\ssep(\tr^R(w), \tr^R(x)) \geq \ssep(w, x).$$
	\end{lemma}
	\begin{proof}
	Let $D = (Q, \Sigma, \delta_D, q_0, F)$ be a DFA that \separates $\tr^R(w)$ and $\tr^R(x)$. 
	We construct a new DFA $E=(Q, \Sigma, \delta_E, q_0, F)$ that \separates $w$ and $x$. For all states $q \in Q$, we set
	$$\delta_E(q, 0) = \delta_D(q, 0), \delta_E(q, 1) = \delta_D(q, 11), \delta_E(q, 2) = \delta_D(q, 01).$$
	It is fairly easy to see that for all strings $u \in \Sigma^*$, we have $\delta_E(u) = \delta_D(\tr^R(u))$.
	Since D separates $\tr^R(w)$ and $\tr^R(x)$, the DFA $E$ separates $w$ and $x$.
\end{proof}
	
	\begin{lemma}
		\label{lem:24}
		Let $t \in \N$ and $R \subseteq \Sets{1,2}^*$ be a regular language such that $\stc(R) \leq t$.
		Also, let $w \in \left(\left(\Sets{1,2}^* - R \right) 0^+ \right)^* (R - \Sets{\epsilon})$.
		For all $w' \in 1\Sets{0,1}^*$, we have
		$$\ssep(\tr(w) f_n w', \tr(w) g_n w') \leq 2t + n + 4.$$
		Recall that $f_n = 0^n$ and  $g_n = 0^{n+(2n+1)!}$.
	\end{lemma}
	\begin{proof}
	We have $\stc(R) \leq t$. So there exists a DFA $D = (Q_D, \Sigma, \delta_D, q_0, F_D) \in M_t$ such that $L(D) = R$.  By using $D$, we construct another DFA $E \in M_{2t+n+4}$ that \distinguishes $\tr(w) f_n w'$ and $\tr(w) g_n w'$. Assume 
	$$Q_D = \Sets{q_0, q_1, \ldots, q_{m-1}}$$
	for some $m \leq t$.
 	We then set $E = (Q, \Sigma, \delta, s, \emptyset)$, where
	$$Q = Q_D \cup \Sets{q'_0, q'_1, \ldots, q'_{m-1}, r_1, \ldots, r_n, r_{n+1}, s, p, p'}.$$
 	Also, we specify the following rules for the transition function of $E$:
	\begin{itemize}
		\item 
		For $0 \leq i \leq m - 1$, we set 
		$$\delta(q_i, 1) = q'_i, \delta(q'_i, 0) = \delta_D(q_i, 2), \delta(q'_i, 1) = \delta_D(q_i, 1).$$
		
		\item
		For all $q_i \notin F_D$, we set $\delta(q_i, 0) = s$.
		\item
		For all $q_i \in F_D$, we set $\delta(q_i, 0) = r_1$. 
	
	\item
	For $1 \leq i \leq n$, we set $\delta(r_i, 0) = r_{i+1}$ and $\delta(r_i,1) = p$.

	\item
	 $\delta(p,0) = \delta(p,1) = p$.
	\item
	$\delta(r_{n+1},0) = r_{n+1}$ and $\delta(r_{n+1},1) = \delta(p',0) = \delta(p',1) = p'$.
	\item
	 $\delta(s, 0) = s$ and $\delta(s, 1) = q'_0$; see Figure \ref{fig:2} for an illustration.
	 \end{itemize}
	
	
	Clearly, for all $0 \leq i \leq m - 1$, we have $\delta(q_i, 11) = \delta_D(q_i, 1)$ and $\delta(q_i, 10) = \delta_D(q_i, 2)$. Hence for all $u \in \Sets{1, 2}^*$, we have $\delta(\tr(u)) = \delta_D(u)$. 
	Since $w \in \left(\left(\Sets{1,2}^* - R \right) 0^+ \right)^* (R - \Sets{\epsilon})$, we have $\delta(\tr(w)) \in F_D$.
	Therefore we can show that $\delta(\tr(w) f_n w') = p$ and $\delta(\tr(w) g_n w') = p'$. Thus $E$ \distinguishes $\tr(w) f_n w'$ and $\tr(w) g_n w'$. So 
	by Lemma \ref{lem:3} 
	we get
	$$\ssep\left(\tr(w) f_n w', \tr(w) g_n w'\right) \leq \size{Q} \leq 2t + n + 4.$$
\begin{figure}
\centering

\includegraphics[width=0.8\textwidth]{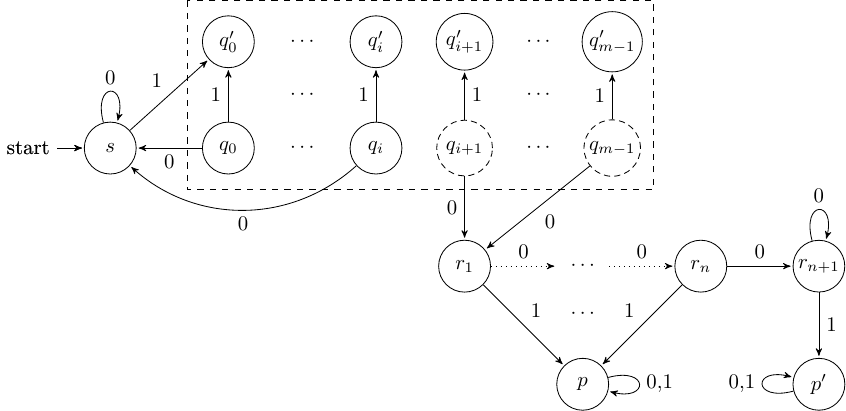}
\caption{The DFA $E$, which is explained in Lemma \ref{lem:24} (assuming $q_{i+1}, \ldots, q_{m-1}$ are the only accept states in $D$).}
  \label{fig:2}
\end{figure}
\end{proof}
	
	\subsection{The Main Result}
	\label{subsec:five}
	Now we are ready to prove our main result. As shown in Theorem \ref{th:final}, by substituting the appropriate values for $n$ and $k$ in Theorem \ref{th:main}, we can prove that the difference $\size{\ssep(w,x)-\ssep(w^R,x^R)}$ is unbounded.

	\begin{theorem}
		\label{th:main}
		For all $k, n \in \N$, there exist two unequal strings $w', x' \in \Sets{0,1}^*$ such that 
		$$\ssep(w', x') \geq \min(2n + 2, 2^{k/2}),$$
		but 
		$$\ssep((w')^R, (x')^R) \leq n + 10k + 10.$$ 
	\end{theorem}
	\begin{proof}
	Let us write $p = \min(2n + 2, 2^{k/2}) - 1$. Consider an arbitrary ordering of all pairs of DFAs in $M_p$ and each of their states:
			$$(D_1, s_1), (D_2, s_2), \ldots, (D_m, s_m),$$
	where $s_i \in Q_{D_i}$, and $m$ is the total number of such pairs, which is clearly finite. Here, for convenience, we use subscript $i$ instead of $D_i$. So let $D_i = (Q_i, \Sigma, \delta_i,q_i, F_i)$. 
	
	We start with 
	$$u_0 = v_0 = z_k, w_0 = u_0 f_n v_0 = z_k f_n z_k, x_0 = u_0 g_n v_0 = z_k g_n z_k.$$
	During the execution of the algorithm that we explain below, we preserve the following conditions for all $0 \leq e \leq m$: 
	\begin{enumerate}
		\item
			\label{cond:first}
			$u_e \in H'_k \left(0^+ H'_k \right)^*$.
		\item
			\label{cond:second}
			$v_e \in z_k \left(0^+ H_k\right)^*$. We have $z_k \in H'_k$ and $H_k \subset H'_k$. Therefore $v_e \in H'_k \left(0^+ H'_k\right)^*$. 
		\item
		\label{cond:third}
			$w_e = u_e f_n v_e$ and $x_e = u_e g_n v_e$.
		\item
			\label{cond:fourth}
			For all $1 \leq j \leq e$ and $\alpha, \alpha' \in \Sigma^*$, if $\delta_j(\alpha u_e) = s_j$, then $\delta_j(\alpha w_e \alpha') = \delta_j(\alpha x_e \alpha')$.
			By setting $\alpha = \alpha' = \epsilon$, it follows that if $\delta_j(u_e) = s_j$ then $\delta_j(w_e) = \delta_j(x_e)$. 
	 \end{enumerate}
	 
	 We can easily observe that these conditions hold for $e = 0$. Now we run the following algorithm iteratively for $i=1, 2, \ldots, m$:
	
	By Condition \ref{cond:third} for $e = i - 1$, we have $w_{i-1}=u_{i-1} f_n v_{i-1}$ and $x_{i-1}=u_{i-1} g_n v_{i-1}$. We 
		set 
		$$u_i = C_n(v_{i-1} 0 u_{i-1}),$$
		where $C_n$ is given by Lemma \ref{lem:9}. By Lemma \ref{lem:9}, we have 
		$$u_i \in v_{i-1} 0 u_{i-1} \left(0^+ v_{i-1} 0 u_{i-1}\right)^*.$$
		So it can be expressed as 
		$$u_i = v_{i-1} 0 u_{i-1} 0^{i_1} v_{i-1} 0 u_{i-1} 0^{i_2} \cdots v_{i-1} 0 u_{i-1} 0^{i_l} v_{i-1} 0 u_{i-1},$$
		for some $l, i_1, \ldots, i_l \in \N$.
		
	By Conditions \ref{cond:first} and \ref{cond:second} for $e=i-1$, we have $u_{i-1}, v_{i-1} \in H'_k \left(0^+ H'_k\right)^*$. Hence by Proposition \ref{prp:five}, we have $v_{i-1} 0 u_{i-1} \in H'_k \left(0^+ H'_k\right)^*$. Therefore by using Proposition \ref{prp:five} again,
	 we get $u_i \in H'_k \left(0^+ H'_k\right)^*$. So Condition \ref{cond:first} holds for $e=i$.
		
	Moreover, let us write
	$$y = u_{i-1} 0^{i_1} v_{i-1} 0 u_{i-1} 0^{i_2} \cdots v_{i-1} 0 u_{i-1} 0^{i_l} v_{i-1} 0 u_{i-1}.$$
	Clearly, we have $u_i = v_{i-1} 0 y$. With the same argument as for $u_i$, by using Proposition \ref{prp:five} we can show that $y \in H'_k \left(0^+ H'_k\right)^*$. 
	We have 
	$\size{Q_i} \leq p \leq {2^{k/2} - 1}$. 	So by applying Proposition \ref{prp:four} to the DFA $D_i$, the states $\delta_i(s_i, f_n v_{i-1} 0)$ and $\delta_i(s_i, g_n v_{i-1} 0)$, and the string $y$, we obtain that $y' \in H_k \left(0^+ H_k \right)^*$ exists such that 
	\begin{equation}
		\label{eq:one}
		\delta_i(s_i, f_n v_{i-1} 0 y) = \delta_i(s_i, f_n v_{i-1} 0 y') 
	\end{equation}	
	and
	\begin{equation}
		\label{eq:two}
		\delta_i(s_i, g_n v_{i-1} 0 y) = \delta_i(s_i, g_n v_{i-1} 0 y').
	\end{equation}
	 Now we set 
	$$v_i = v_{i-1} 0 y'.$$
	By Condition \ref{cond:second} for $e=i-1$, we have $v_{i-1} \in z_k \left(0^+ H_k\right)^*$. 
	Thus we obtain  
	$$v_i \in z_k \left(0^+ H_k\right)^* 0^+ H_k \left(0^+ H_k \right)^* = z_k \left(0^+ H_k\right)^+ \subseteq z_k \left(0^+ H_k\right)^*,$$
	and therefore Condition \ref{cond:second} is satisfied for $e = i$.
	Afterwards, we set $w_i \coloneqq u_i f_n v_i$ and $x_i \coloneqq u_i g_n v_i$. This satisfies Condition \ref{cond:third} for $e=i$. 
	
	By substituting $u_i = v_{i-1} 0 y$ and $v_i = v_{i-1} 0 y'$ in equations \ref{eq:one} and \ref{eq:two}, we get
	$\delta_i(s_i, f_n u_i) = \delta_i(s_i, f_n v_i)$ and $\delta_i(s_i, g_n u_i) = \delta_i(s_i, g_n v_i)$.
	Now consider an arbitrary string $\alpha \in \Sigma^*$. Suppose $\delta_i(\alpha u_i)=s_i$. Hence we get
	\begin{equation}
		\label{eq:three}
		\delta_i(\alpha u_i f_n u_i) = \delta_i(s_i, f_n u_i) = \delta_i(s_i, f_n v_i) = \delta_i(\alpha u_i f_n v_i),
	\end{equation}
	and similarly, we have
	\begin{equation}
		 \label{eq:four}
		 \delta_i(\alpha u_i g_n u_i) = \delta_i(s_i, g_n u_i) = \delta_i(s_i, g_n v_i) = \delta_i(\alpha u_i g_n v_i).
	\end{equation}
	Besides, $u_i = C_n(v_{i-1} 0 u_{i-1})$. So by Lemma \ref{lem:9}, we get that $\ssep(u_i f_n u_i, u_i g_n u_i)$ is at least $2n+2$. Hence by Lemma \ref{lem:4}, we get
	$$\ssep(\alpha u_i f_n u_i,  \alpha u_i g_n u_i) \geq \ssep(u_i f_n u_i, u_i g_n u_i) \geq 2n+2.$$
	Since $p \leq 2n+1$, no $D$ in $M_p$ can \separate $\alpha u_i f_n u_i$ and $\alpha u_i g_n u_i$. 
	Hence by Lemma \ref{lem:3}, $D_i$ cannot \distinguish $\alpha u_i f_n u_i$ and $\alpha u_i g_n u_i$,
	so we have 
	\begin{equation}
		\label{eq:five}
		\delta_i(\alpha u_i f_n u_i) = \delta_i(\alpha u_i g_n u_i).
	\end{equation}
	By equations \ref{eq:three}, \ref{eq:four}, and \ref{eq:five}, we can conclude that if $\delta_i(\alpha u_i)=s_i$, then we have
	$$\delta_i(\alpha u_i f_n v_i) = \delta_i(\alpha u_i f_n u_i) = \delta_i(\alpha u_i g_n u_i) = \delta_i(\alpha u_i g_n v_i),$$
	or equivalently, by substituting $w_i = u_i f_n v_i$ and $x_i = u_i g_n v_i$, we can write $\delta_i(\alpha w_i) = \delta_i(\alpha x_i)$. Furthermore,
	it follows that for any string $\alpha' \in \Sigma^*$, we have 
	$$\delta_i(\alpha w_i \alpha') = \delta_i(\alpha x_i \alpha').$$
	Thus Condition \ref{cond:fourth} is satisfied when $e = j = i$. Moreover, we have $u_i \in \Sigma^*u_{i-1}$ and $v_i \in v_{i-1} \Sigma^*$. So there exist $b, b' \in \Sigma^*$ such that $u_i = b u_{i - 1}$ and $v_i = v_{i - 1} b'$. Hence we can write
	$$w_i = u_i f_n v_i = b u_{i - 1} f_n v_{i - 1} b' = b w_{i-1} b',$$
	and similarly, we have
	$$x_i = u_i g_n v_i = b u_{i - 1} g_n v_{i - 1} b' = b x_{i-1} b'.$$
	Thus for all $ \alpha, \alpha' \in \Sigma^*$, we have
	\begin{equation}
		\label{eq:six}
		\alpha w_i \alpha' =  \alpha b w_{i-1} b'  \alpha',  \alpha x_i \alpha' =  \alpha b x_{i-1} b'  \alpha'.
	\end{equation}
	Suppose $1 \leq j \leq i - 1$. By Condition \ref{cond:fourth} for $e = i - 1$, if 
	$$\delta_j(\alpha b u_{i - 1}) = \delta_j(\alpha u_i) = s_j,$$ 
	then $\delta_j(\alpha b w_{i-1} b' \alpha') = \delta_j(\alpha b x_{i-1} b' \alpha')$, or equivalently, by using equation \ref{eq:six}
	we can write $\delta_j(\alpha w_i \alpha') = \delta_j(\alpha x_i \alpha')$.
	So Condition \ref{cond:fourth} for $e=i$ is also satisfied when $j \leq i - 1$. Therefore Condition \ref{cond:fourth} holds for $e=i$. Hence we proved that all four conditions are satisfied for $e = i$.
		
	In the end, we set $u=u_m, v=v_m, w=w_m$ and $x=x_m$. We claim $\ssep(w, x) \geq p + 1$. Otherwise, suppose $D \in M_p$ \separates $w,x$. So $D$ \distinguishes $w$ and $x$. Let us write $s = \delta_D(u)$. We have $\size{Q_D} \leq p$. Therefore by definition, there exists $1 \leq i \leq m$ such that $D_i = D$ and $s_i = s$. Since $\delta_D(u) = s$, by Condition \ref{cond:fourth} for $e = m$ and $j = i$, we have $\delta_D(w) = \delta_D(x)$, which contradicts the assumption that $D$ \separates $w$ and $x$. Therefore $\ssep(w,x) \geq p + 1$. Now we set $w' = \tr^R(w)$ and $x' = \tr^R(x)$. By Lemma \ref{lem:23}, we have
		$$\ssep(w', x') \geq \ssep(w, x) \geq p + 1 = \min(2n + 2, 2^{k/2}).$$
	Besides, we have $w^R=v^R f_n u^R$ and $x^R=v^R g_n u^R$. By Condition \ref{cond:second}, we have $v \in z_k \left(0^+ H_k\right)^*$. Thus  
	
	\begin{equation*}
  		\begin{split}
  		v^R \in \left(H_k^R 0^+\right)^* z_k^R &= \left( \left(\Sets{1,2}^*-G_k^R \right) 0^+\right)^* z_k^R \\ 
  			&\subseteq \left( \left(\Sets{1,2}^*-G_k^R \right) 0^+\right)^* (G_k^R - \Sets{\epsilon}).
		\end{split}
	\end{equation*}
	By Lemma \ref{lem:14}, we have $\stc(G_k^R) \leq 5k + 3$. Moreover, by Condition \ref{cond:first} we obtain $u^R \in \Sets{1,2}\Sigma^*$. Therefore by definition, we obtain $\tr(u^R) \in 1\Sets{0,1}^*$. Hence 
	by Lemma \ref{lem:24}, we get
	\begin{equation*}
		\begin{aligned}
			\ssep((w')^R, (x')^R) 
			&= \ssep(\tr(w^R),\tr(x^R)) \\
			&= \ssep(\tr(v^R f_n u^R), \tr(v^R g_n u^R)) \\
			&= \ssep(\tr(v^R) f_n \tr(u^R), \tr(v^R) g_n \tr(u^R)) && \leq 2(5k + 3) + n + 4			\\
			& &&= 10k + n + 10.
		\end{aligned}	
	\end{equation*}
\end{proof}

	\begin{theorem}
		\label{th:final}
		The difference 
		$$\size{\ssep(w,x)-\ssep(w^R,x^R)}$$ is unbounded for an alphabet of size at least $2$. 
	\end{theorem}
	\begin{proof}
		Let 
		 $k$ be a positive even integer. If we set $n = 2^{k/2 - 1} - 1$, then by Theorem \ref{th:main}, there exist strings $w$ and $x$ in $\Sets{0,1}^*$ such that 
		$$\ssep(w, x) \geq \min(2n+2, 2^{k/2}) = 2^{k/2},$$
		and
		$$ \ssep(w^R, x^R) \leq n + 10k + 10 = (2^{k/2 - 1} - 1) + 10k + 10.$$
		So we have
		\begin{equation*}
  			\begin{split}
  				\ssep(w, x) - \ssep(w^R, x^R) & \geq 2^{k/2} - \left( 2^{k/2 - 1} + 10k + 9 \right) \\
  				& = 2^{k/2 - 1} - 10k - 9,
			\end{split}
		\end{equation*}
		which tends to infinity as $k$ approaches infinity. 
\end{proof}

	\section{Conclusion}
	In this paper, we proved that the difference $\size{\ssep(w,x) - \ssep(w^R,x^R)}$ can be unbounded. However, it remains open to determine whether there is a good upper bound on $\ssep(w,x) / \ssep(w^R,x^R)$.

	\section*{Acknowledgments}
I wish to thank Jeffrey Shallit, Mohammad Izadi, Arseny Shur, MohammadTaghi Hajiaghayi, Keivan Alizadeh, Hooman Hashemi, Hadi Khodabandeh, and Mobin Yahyazadeh, who helped me write this paper. I would also like to thank the anonymous referees for their careful reading of this paper, and for their valuable comments and suggestions. 
	\bibliography{references/references}
\end{document}